\documentclass[twocolumn,english,aps,showpacs,superscriptaddress]{revtex4-1}
\usepackage[latin9]{inputenc}
\usepackage{amsmath}
\usepackage{amssymb}
\usepackage{mathdots}
\usepackage{graphicx}
\usepackage{esint}
\usepackage{float}

\makeatletter

\newcommand{\lyxdot}{.}

\newcommand{\e}{\mathrm{e}}

\makeatother

\usepackage{babel}
\begin{document}

\title{Genesis of the Floquet Hofstadter butterfly}

\author{S. H. Kooi}

\affiliation{Institute for Theoretical Physics, Center for Extreme Matter and
Emergent Phenomena,\\ Utrecht University, Princetonplein 5, 3584
CC Utrecht, the Netherlands}

\author{A. Quelle}

\affiliation{Institute for Theoretical Physics, Center for Extreme Matter and
Emergent Phenomena,\\ Utrecht University, Princetonplein 5, 3584
CC Utrecht, the Netherlands}

\author{W. Beugeling}

\affiliation{Physikalisches Institut, Universität Würzburg, Am Hubland, 97074
Würzburg, Germany}

\author{C. Morais Smith}

\affiliation{Institute for Theoretical Physics, Center for Extreme Matter and
Emergent Phenomena,\\ Utrecht University, Princetonplein 5, 3584
CC Utrecht, the Netherlands}

\date{\today} \begin{abstract} We investigate theoretically the spectrum
of a graphene-like sample (honeycomb lattice) subjected to a perpendicular
magnetic field and irradiated by circularly polarized light. This system is
studied using the Floquet formalism, and the resulting Hofstadter spectrum is
analyzed for different regimes of the driving frequency. For lower frequencies,
resonances of various copies of the spectrum lead to intricate formations of
topological gaps. In the Landau-level regime, new wing-like gaps emerge upon
reducing the driving frequency, thus revealing the possibility of dynamically
tuning the formation of the Hofstadter butterfly. In this regime, an effective
model may be analytically derived, which allows us to retrace the energy levels
that exhibit avoided crossings and ultimately lead to gap structures with a
wing-like shape. At high frequencies, we find that gaps open for various fluxes
at $E=0$, and upon increasing the amplitude of the driving, gaps also close and
reopen at other energies. The topological invariants of these gaps are
calculated and the resulting spectrum is elucidated. We suggest opportunities
for experimental realization and discuss similarities with Landau-level
structures in non-driven systems. \end{abstract} \maketitle

\section{Introduction}

The complex fractal structure of the Hofstadter butterfly, which reveals the
interplay between the lattice constant and the magnetic length when a
perpendicular magnetic field is applied to a crystal lattice, has fascinated
researchers since its first theoretical prediction
\cite{hofstadter1976energy}. However, its experimental realization seemed to
be impossible at first sight, because for typical crystal lattice spacings,
the magnetic field required to observe the butterfly is of the order of
thousands of tesla. Recently, moiré superlattices, obtained when depositing
graphene on mismatched substrates, such as hBN, have been realized
\cite{yankowitz2012emergencehBn,xue2011scanninghBn}. These structures have an
effective lattice spacing that is an order of magnitude larger than the usual
crystal lattices. This has brought the required magnetic-field strength within
experimental reach, and enabled the observation of the Hofstadter butterfly
spectra \cite{GeimExperiment,Kimexperiment}. In addition, the Hofstadter
butterfly has been proposed in nanophotonic devices
\cite{nanophotonicHofstadter}, and for bosons in optical lattices
\cite{ZollerBosonHofstadter,dalibard2011colloquium,goldman2014light}, where it
has also been experimentally realized \cite{KetterleBosonHofstadter,BlochBosonHofstadter}.

All these studies were done in equilibrium, and so far out-of-equilibrium
Hofstadter setups have not received much attention, although driven systems
have been under intense scrutiny recently \cite{eckardt2017colloquium,titum2016anomalousFloquetPump,lindner2011floquet,zheng2014floquetShakingFloquet,quelle2014dynamical,delplace2013mergingConesGraphene,koghee2012merging,benito2014floquet,rechtsman2013photonic,wang2013observation,flaschner2016experimental,WangGong2008,HoGong2014,kickedHarper,HarmonicDrivenHarper,ding2018quantum}. In particular, time-periodic
driving attracted great interest because it can be conveniently described in
the framework of Floquet theory
\cite{lindner2011floquet,eckardt2017colloquium,goldman2015periodically,gomez2013floquetBlockexpansion}. This
allows one to define quasi-static properties of the driven system that can be
measured, and is a tuning knob for quantum simulations both in condensed-matter and cold-atom experiments. The quasi-energy spectrum obtained using
Floquet theory is periodic, with a period proportional to the driving 
frequency. Recently, periodically driven systems have been observed in
photonics \cite{rechtsman2013photonic}, condensed-matter
\cite{wang2013observation}, and cold-atom experiments
\cite{flaschner2016experimental,jotzu2014experimental}.

Periodic driving described by Floquet theory can lead to many interesting
topological phase transitions \cite{titum2016anomalousFloquetPump,lindner2011floquet,zheng2014floquetShakingFloquet,quelle2014dynamical,delplace2013mergingConesGraphene,koghee2012merging,MikamiEA2016}, characterized by a slightly different
topological invariant than for the undriven case \cite{rudner2013anomalous,kitagawa2010topological,nathan2015topologicalUnifiedClassication,quelle2016bandwidth}. For example, Floquet theory predicts additional topological phases in
the Kitaev chain \cite{benito2014floquet}. Topological behavior induced by
periodic driving has been observed experimentally in photonic waveguides
\cite{rechtsman2013photonic}, and a gap opening has been detected on the
surface of a topological insulator upon irradiation with circularly polarized
light \cite{wang2013observation}. The Berry curvature of such Floquet Bloch
bands has also been explicitly measured \cite{flaschner2016experimental}.

\begin{figure}[!b]
\includegraphics[width=8cm]{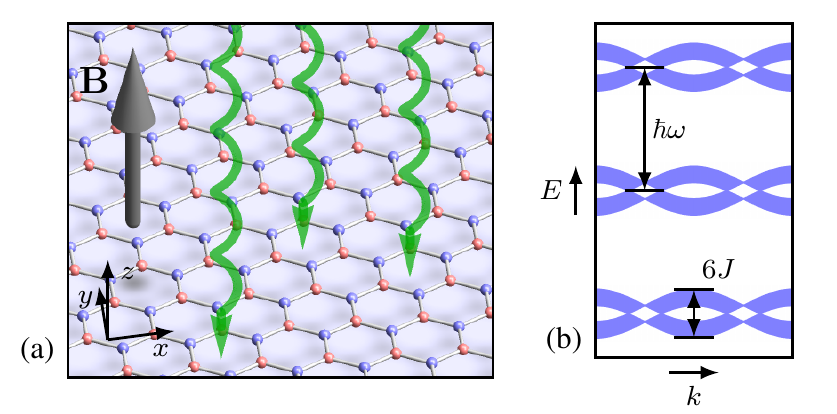}

\caption{(a) Schematic setup of our model. The honeycomb lattice is subjected
to a perpendicular magnetic field, and is simultaneously irradiated by
circularly polarized light. (b) Quasi-energy spectrum of the Floquet model. The generic
feature of the Floquet spectrum is the periodicity with $\hbar\omega$ in the
vertical direction. \label{fig:setup}} \end{figure}

Several recent works have been dedicated to the investigation of the driven Hofstadter model.
In Refs.~\cite{kickedHarper,HarmonicDrivenHarper}, the driven Hofstadter 
model has been investigated on a square lattice for a specific flux $\phi=1/3$ (in units of
the flux quantum $\phi_0=h/e$), and for two different driving protocols. In
both cases, the authors find counter-propagating edge modes in the quasi-energy spectrum,
crossing $E=\pm \pi\hbar/T$, where $T$ is the period of the driving. The Hofstadter butterfly for a driven honeycomb lattice
has been studied in Ref.~\cite{wackerl2018driven}, with an extensive Chern number analysis.
In Ref.~\cite{ding2018quantum}, a
transition from the half-integer to the integer quantum Hall effect has been
theoretically proposed to occur upon elliptical driving of an ac field.

Here, we show that the Floquet method can be used to unveil the formation of
the Hofstadter butterfly at low magnetic fields by adding a periodic driving.
Upon tuning the frequency, the bands start to overlap and avoided crossings
occur, that lead to the formation of wings. At small flux, where the spectrum
has a Landau-level structure, the procedure can be analytically monitored
using the Floquet formalism. In doing so, we gain insight on the mechanism of hybridization between Landau levels. For larger magnetic fields, we perform numerical
calculations to obtain the full butterfly spectrum for various frequencies.

\begin{figure*}[!t]
\includegraphics[width=17cm]{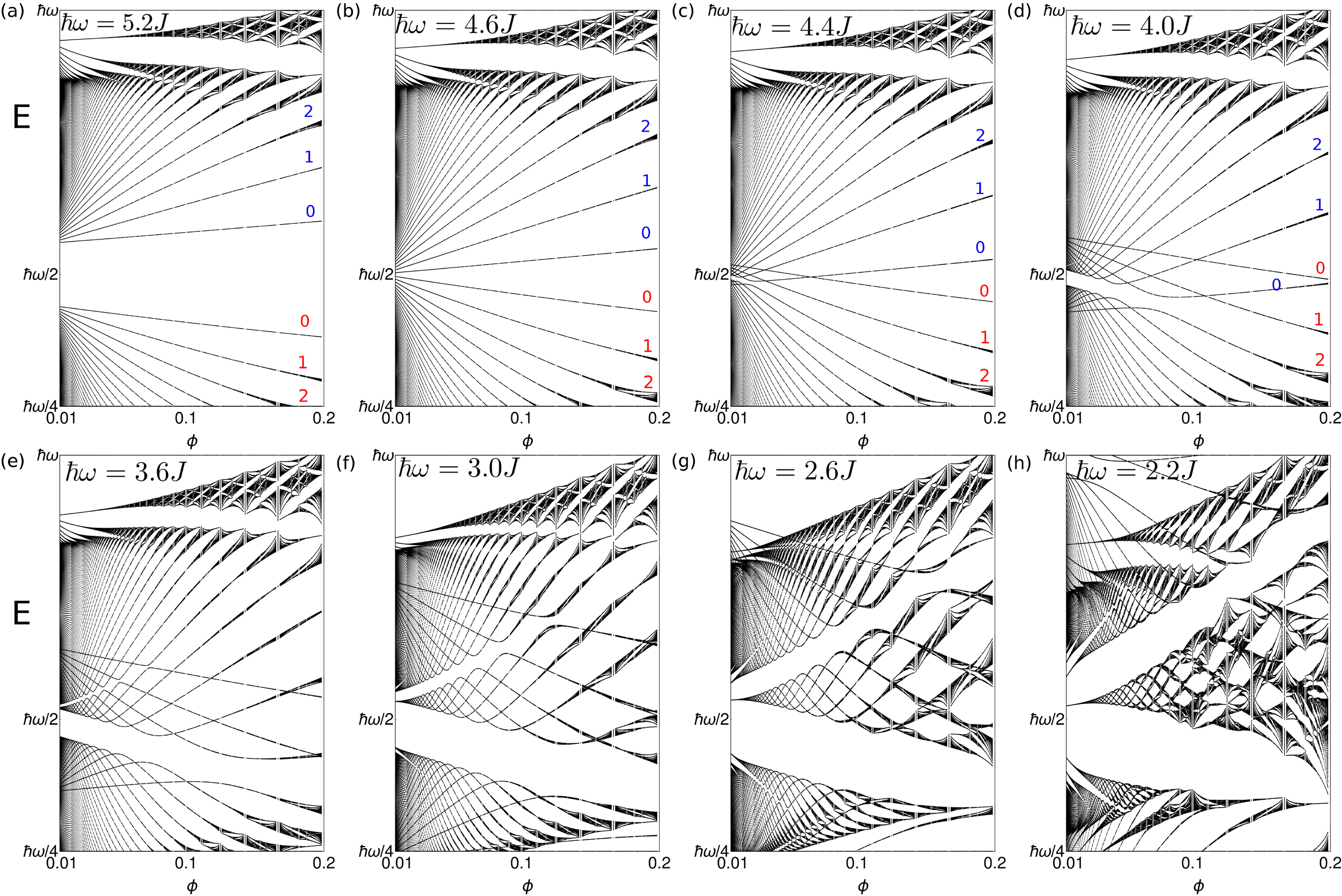}

\caption{Spectra for small fluxes $\phi$, in the Landau-level regime, plotted for various
values of $\omega$, with $A_{0}=1$. In the first four plots we label 
the first three Landau levels of the upper (blue) and lower (red) copy.
\label{fig:-1}}
\end{figure*}

The outline of this paper is as follows. In Sec.~\ref{sec:Model},
we introduce the model and explain the details of Floquet theory.
In Sec.~\ref{sec:High-frequency-approximation}, we present numerical
results for the small-flux regime and derive an effective model to
explain the mixing of the Landau levels. In Sec.~\ref{sec:Numerical-results},
we present and analyze our numerical results for the full range of flux,
in both high- and low-frequency regimes. Our findings are summarized
in Sec.~\ref{sec:Conclusion}.

\section{The model\label{sec:Model}}

We consider a honeycomb lattice (e.g., a graphene monolayer) subject to a
perpendicular magnetic field
and to irradiation by circularly polarized light [see Fig.~\ref{fig:setup}(a)].
The system is described by a tight-binding model of electrons on a
honeycomb lattice, where the background magnetic field and the circularly
polarized light are included through a vector potential $\mathbf{A}$,
via Peierls substitution. The Hamiltonian reads
\begin{align}
H & =-J\sum_{l=1,2,3}\sum_{\mathbf{r}}|\mathbf{r}+\mathbf{\delta}_{l}\rangle\e^{i\int d\mathbf{s}\cdot\mathbf{A}}\langle\mathbf{r}|+\mathrm{H.c.},\label{eq:}
\end{align}
where $J$ is the hopping parameter, $\mathbf{r}$ is the position
of a site, $\delta_{l}$ are the nearest-neighbor vectors of the honeycomb
lattice, and $d\mathbf{s}$ parametrizes the path between two sites
$\mathbf{r}$ and $\mathbf{r}+\delta_{l}$. The vector potential consists
of two contributions,
\begin{align}
\mathbf{A}(\mathbf{r},t) & =\mathbf{A}_{\mathrm{mag}}(\mathbf{r})+\mathbf{A}_{\mathrm{light}}(t).\label{eq:-1}
\end{align}
 The first term is due to the background magnetic field, which will
be described in the Landau gauge, 
\begin{align}
\mathbf{A}_{\mathrm{mag}}(\mathbf{r}) & =-\frac{e}{\hbar}(By,\,0),\label{eq:-2}
\end{align}
where $e$ is the electron charge and $\hbar$ the reduced Planck
constant. The second contribution comes from the circularly polarized
light and is periodic in time, 
\begin{align}
\mathbf{A}_{\mathrm{light}}(t) & =A_{0}\left(-\sin\left(\omega t\right),\cos\left(\omega t\right)\right),\label{eq:l}
\end{align}
where $A_{0}$ is the amplitude, $\omega$ is the frequency of the
light and $t$ denotes time.

Let us start by considering the static Hamiltonian, with $\mathbf{A}_{\mathrm{light}}=0$. A Fourier transformation then yields the Harper equation of the honeycomb
lattice (we set the lattice spacing to unity):
\begin{align}
 & -E\left(\begin{array}{cc}
1 & 0\\
0 & 1
\end{array}\right)\psi_{r}\left(k\right)=J\left(\begin{array}{cc}
0 & 1\\
1 & 0
\end{array}\right)\psi_{r}\left(k\right)\nonumber \\
+J & e^{i\frac{\sqrt{3}}{2}k_{x}}\left(\begin{array}{cc}
0 & e^{\frac{3}{2}i\left(k_{y}+\frac{2}{3}\pi\frac{r\,p}{q}\right)}\\
e^{-\frac{3}{2}i\left(k_{y}+\frac{2}{3}\pi\frac{r\,p}{q}\right)} & 0
\end{array}\right)\psi_{r+1}\left(k\right)\nonumber \\
+J & e^{-i\frac{\sqrt{3}}{2}k_{x}}\left(\begin{array}{cc}
0 & e^{\frac{3}{2}i\left(k_{y}+\frac{2}{3}\pi\frac{r\,p}{q}\right)}\\
e^{-\frac{3}{2}i\left(k_{y}+\frac{2}{3}\pi\frac{r\,p}{q}\right)} & 0
\end{array}\right)\psi_{r-1}\left(k\right),\label{eq:-3}
\end{align}
where $r=1,\ldots,q$, and $\psi_{q+1}=\psi_{1}$, with
\begin{align}
 \psi_{r}(k)
  & = \left( \psi^\mathrm{A}_{r,k_{x},k_{y}}, \psi^\mathrm{B}_{r,k_{x},k_{y}} \right).
\end{align}
Here, the components $\psi^\mathrm{A}$ and $\psi^\mathrm{B}$ refer to the two sublattices
of the honeycomb lattice, and we have taken the flux per unit cell
to be $\phi=p/q$ in units of the flux quantum $\phi_0$, where $p$
and $q$ are co-prime integers. Thus, the matrices in the Harper equation [Eq.~\eqref{eq:-3}] act in sublattice space.

To describe the influence of the circularly polarized light, we now
also consider $\mathbf{A}_{\mathrm{light}}$. This will amount to
each hopping picking up a phase, 
\begin{align}
\e^{i\int d\mathbf{s}\cdot\mathbf{\mathbf{A}_{\mathrm{light}}}}= & \exp\left\{ iA_{0}\left[-\cos\left(\theta\right)\sin\left(\omega t\right)+\sin\left(\theta\right)\cos\left(\omega t\right)\right]\right\} ,\label{eq:-4}
\end{align}
where $\theta$ is the angle between the bond and the $x$-axis.
Since the Hamiltonian is now periodic in time, we can define the Floquet
Hamiltonian by \cite{eckardt2017colloquium} 
\begin{align}
H_{\mathrm{Floq}} & =\frac{i\hbar}{T}\ln\left[U\left(T,0\right)\right].\label{eq:-13}
\end{align}
Here, $T=2\pi/\omega$ is the period of the driving and $U\left(T,0\right)$ is
the time-evolution operator, which may be found by numerically solving
the Schrödinger equation
\begin{align}
i\hbar\frac{\partial U\left(t,t'\right)}{\partial t} & =H\left(t\right)\,U\left(t,t'\right).\label{eq:-5}
\end{align}
By calculating the eigenvalues and eigenstates of $H_{\mathrm{Floq}}$,
we can determine the quasi-static behavior of the system at stroboscopic
timescales larger than $T$. The time-periodic Hamiltonian can thus be expanded
into the Fourier coefficients $H_n$, as
\begin{align}
H(t) & =\sum_{n}H_{n}\e^{in\omega t}.\label{eq:-6}
\end{align}
The eigenenergies of the Floquet Hamiltonian then follow from diagonalization of
\begin{align}
H_{\mathrm{Floq}}= & \left(\begin{array}{ccccc}
\ddots & \vdots & \vdots & \vdots & \iddots\\
\cdots & H_{0}+\hbar\omega & H_{1} & H_{2} & \cdots\\
\cdots & H_{-1} & H_{0} & H_{1} & \cdots\\
\cdots & H_{-2} & H_{-1} & H_{0}-\hbar\omega & \cdots\\
\iddots & \vdots & \vdots & \vdots & \ddots
\end{array}\right).\label{eq:-8}
\end{align}
We can interpret the Hamiltonian Eq.~\eqref{eq:-8} as an infinite
set of copies of the Hamiltonian $H_{0}$, separated by energies $\hbar\omega$,
as illustrated by Fig.~\ref{fig:setup}(b).
These copies are then mixed by the off-diagonal elements. If $\hbar\omega$
is much larger than the bandwidth $6J$ of the spectrum of $H_{0}$,
this mixing will be negligible. However, when $\hbar\omega$ becomes comparable
to $6J$, the different copies of $H_{0}$ start to overlap and
the mixing terms become important.

For high frequencies,
the Floquet Hamiltonian can be expanded to first non-trivial order as
\cite{grozdanov1988quantum,eckardt2017colloquium,LopezEA2015,eckardt2015high,rahav2003effective,MikamiEA2016}
\begin{equation}\label{eq:Floquetpert}
  H_\mathrm{Floq} \approx H_0 + [H_{-1}, H_{1}]/\hbar \omega.
\end{equation}

\section{Landau-level regime\label{sec:High-frequency-approximation}}

We will first focus our attention to the small-flux limit, where the
Hofstadter spectrum typically exhibits a Landau-level structure.
In Fig.~\ref{fig:-1}, we plot the energies as a function
of the flux $\phi$ for different values of the driving frequency $\omega$.
We consider the regime where the frequency becomes comparable to the
bandwidth $6J$. In Fig.~\ref{fig:-1}(a), for $\hbar\omega=5.2J$,
we observe that two subsequent copies of $H_{0}$ are still well separated.
(The two copies shown here live in the intervals $[-\hbar\omega/2,\hbar\omega/2]$ and $[\hbar\omega/2,3\hbar\omega/2]$, respectively.)
The coupling between the two copies reduces their width to a value smaller than $6J$.
Upon lowering the frequency, the two
copies of bands come closer to each other and start to overlap.
We see this process in Figs.~\ref{fig:-1}(c)-(h). In Figs.~\ref{fig:-1}(b)
and \ref{fig:-1}(c), the initial overlap of the bands takes place.
Curiously, the top two Landau levels of the lower copy do not mix with any
Landau level of the upper copy, while the rest hybridizes and a gap opens due
to their avoided crossings. We will explain this behavior in the next
section using an effective model to describe this regime. In Figs.~\ref{fig:-1}(e)-(h),
one sees additional gaps opening, and one observes that the two largest
gaps acquire a shape that resembles the wings of the undriven Hofstadter
butterfly.

\begin{figure*}[!t]
\includegraphics[width=8.5cm]{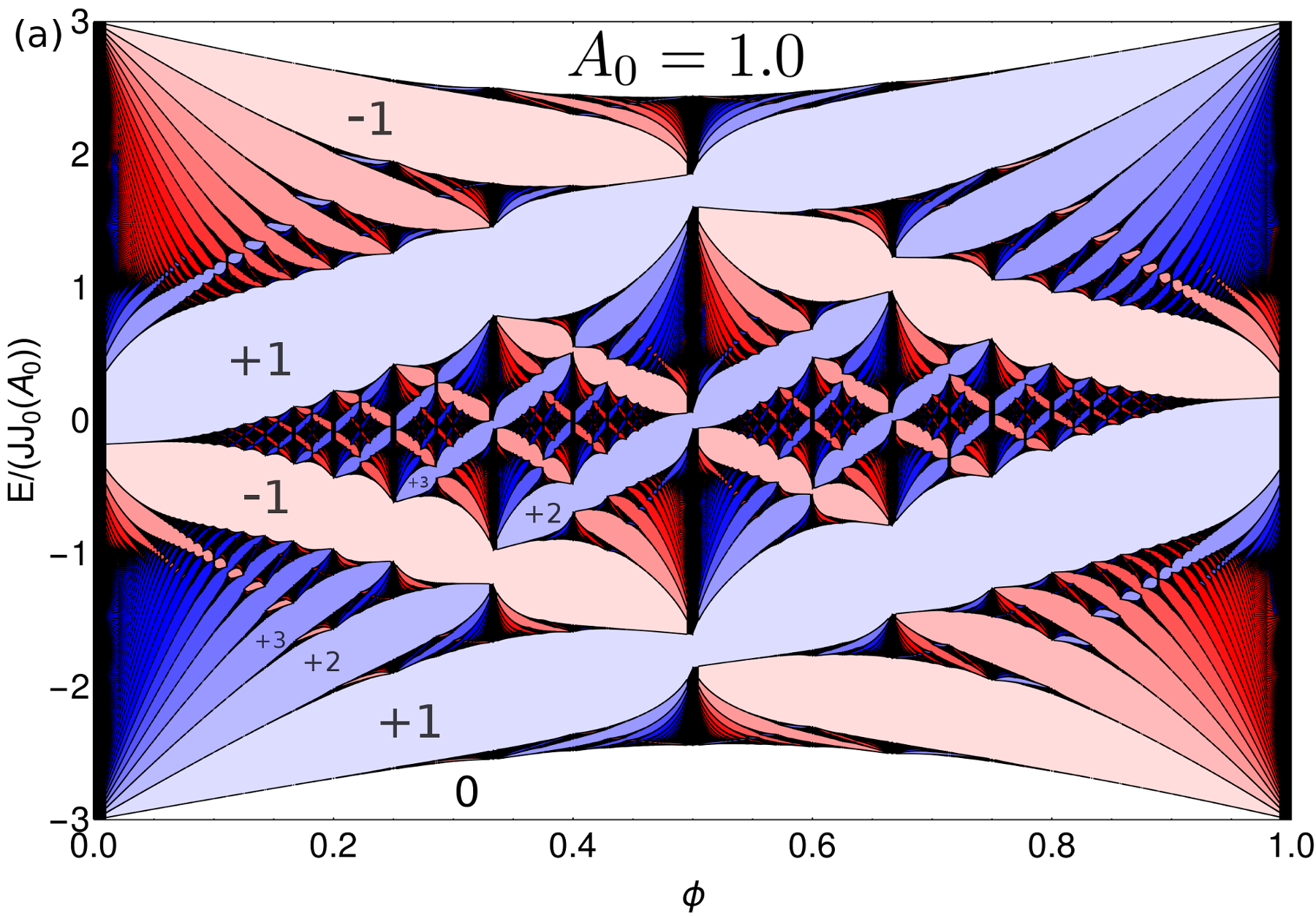}\includegraphics[width=8.5cm]{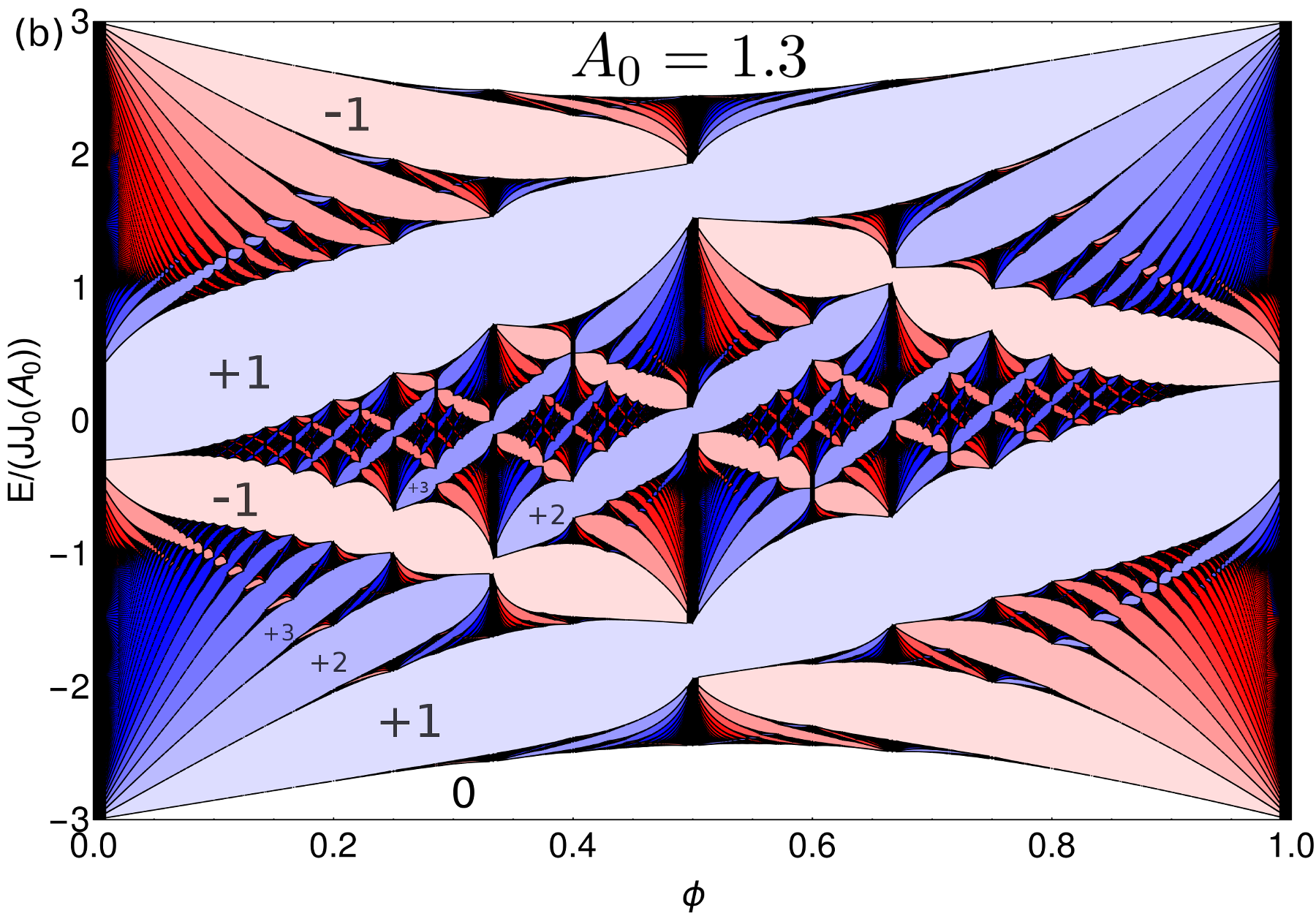}

\includegraphics[width=8.5cm]{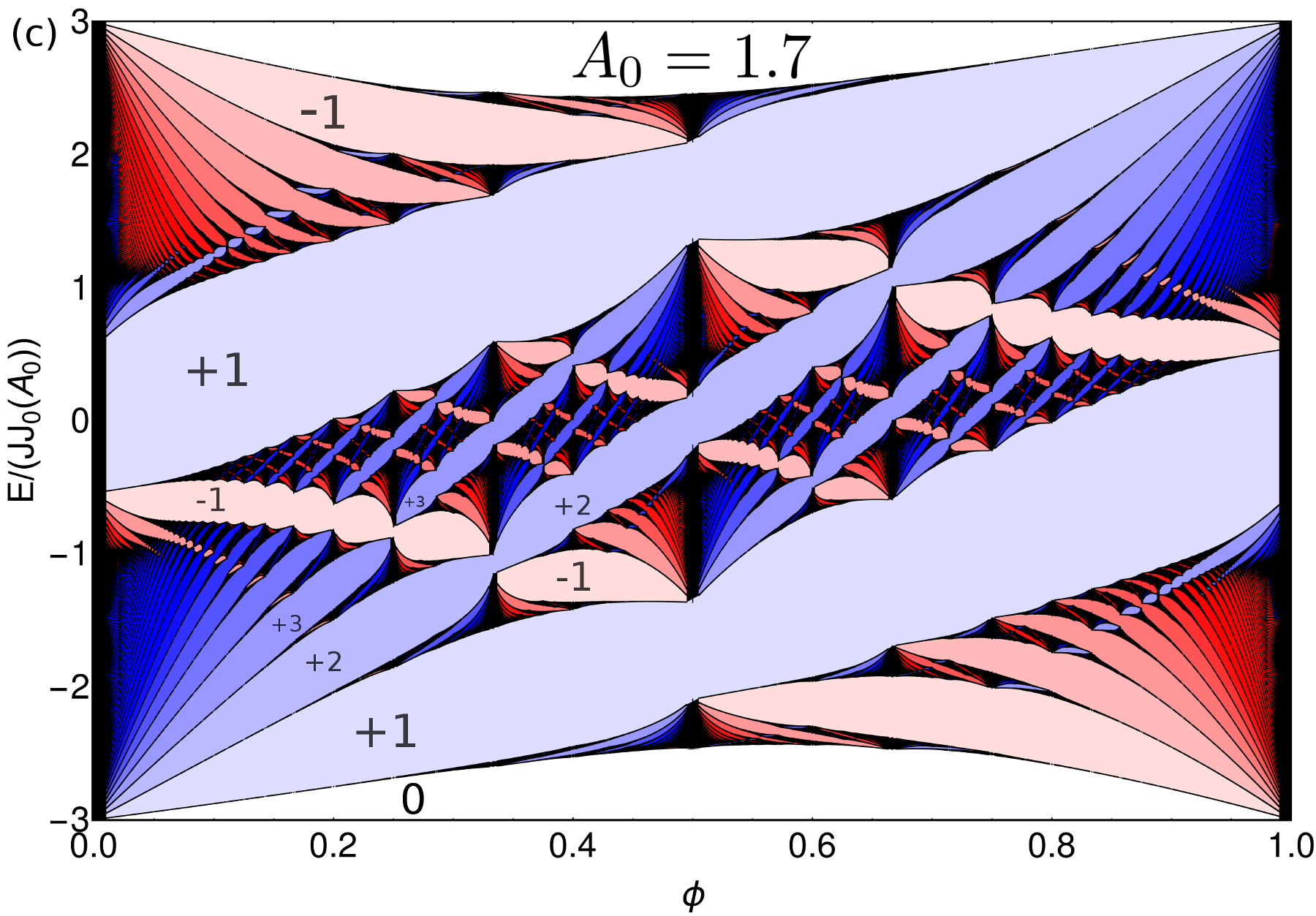}\includegraphics[width=8.5cm]{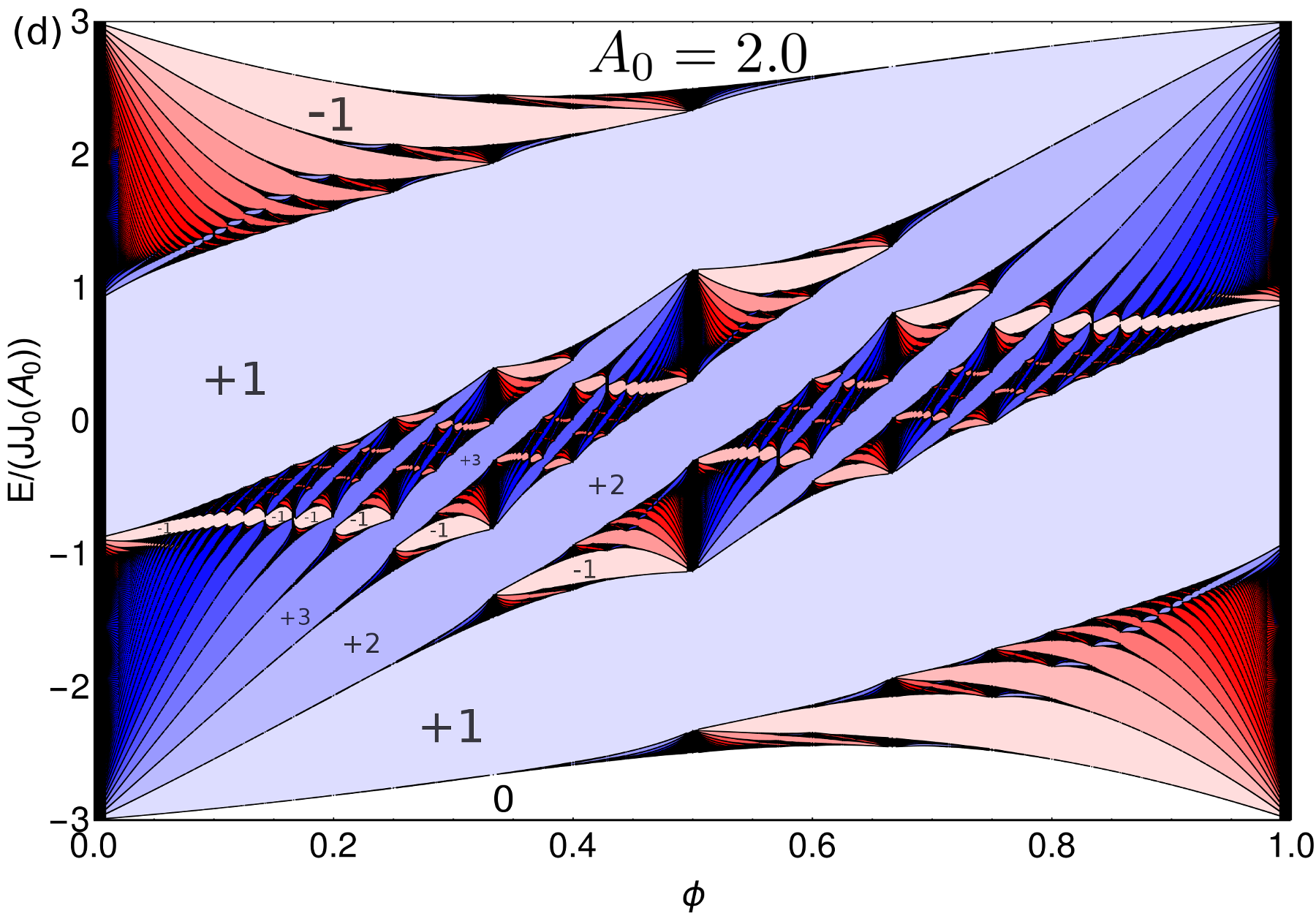}

\includegraphics[width=15cm]{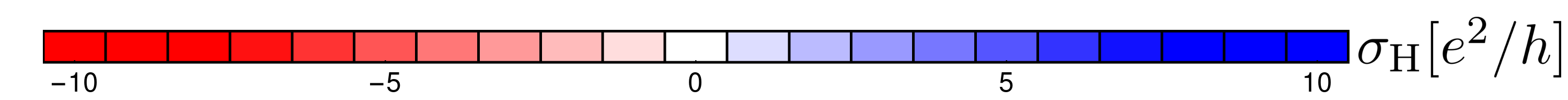}

\caption{Full spectrum plotted for $\hbar\omega=12J$ and various values of
$A_{0}$. The colors of the gaps correspond to the number of left
(red) or right (blue) moving edge states.\label{fig:Butterfly-spectrum-for}}
\end{figure*}

\subsection*{Effective model}

We now derive an effective model to describe the initial overlap of
the two copies of bands displayed in Figs.~\ref{fig:-1}(b)-(d),
aiming at understanding why the top two bands of the lower copy do
not hybridize with the bands of the upper copy. To do so, we zoom
in around $E\approx\hbar\omega/2$, where the overlap occurs. Our starting
point is the Hamiltonian in Eq.~\eqref{eq:-8}. Since we are interested
in the regime where two copies start overlapping, $\hbar\omega\lesssim6J$,
at energy $E\approx\hbar\omega/2$, we can restrict ourselves to two
copies of $H_{0}$.
Here, we take the ones centered at $E=0$ and $E=\hbar\omega$,
and consider their mixing, of which the dominant contribution stems from
$H_{1}$ and $H_{-1}$. The mixing with levels in more distant Floquet copies is negligible,
as the effect is suppressed with increasing energy difference. (A similar analysis is done
in Ref.~\cite{quelle2016bandwidth}.)
The effective Hamiltonian then becomes
\begin{align}
H_{\mathrm{Floq}}^{\mathrm{eff}} & =\begin{pmatrix}H_{0}+\hbar\omega & H_{1}\\
H_{-1} & H_{0}
\end{pmatrix}.\label{eq:-10}
\end{align}

To derive analytical expressions for $H_{n}$, with $n=-1,0,1$,
we initially solve the problem at zero dc magnetic field ($\phi=0$),
including only the time-dependent
perturbation. In this case, $H_{n}$ can be obtained by making the following
substitution in the Hamiltonian,
\begin{align}
J & \rightarrow\frac{J}{T}\int_{0}^{T}dt' \exp\left\{iA_{0}\left[-\cos\left(\theta\right)\sin\left(\omega t'\right)\right.\right.\nonumber \\
 & +\sin\left(\theta\right)\cos\left(\omega t'\right)\Bigr]\Bigr\}\exp\left\{ in\omega t'\right\} \nonumber \\
 & =J\,J_{n}\left(A_{0}\right)\exp\left[in\left(\theta+\frac{\pi}{2}\right)\right],\label{eq:-7}
\end{align}
where $J_{n}$ is the $n\mathrm{th}$ Bessel function of the first kind.
Applying this substitution to the tight-binding Hamiltonian [Eq.~\eqref{eq:}]
of the honeycomb lattice, we obtain
\begin{equation}
  H_n = 
  \begin{pmatrix}
    0 & h_n\\h'_n & 0
  \end{pmatrix},
\end{equation}
where
\begin{align}
h_n  & =
-J\left(\e^{i\left(k_{y}\frac{1}{2}+k_{x}\frac{\sqrt{3}}{2}\right)}\e^{in\frac{5}{6}\pi}\right.\nonumber \\
 & \left.+\e^{i\left(k_{y}\frac{1}{2}-k_{x}\frac{\sqrt{3}}{2}\right)}\e^{in\frac{\pi}{6}}+\e^{-ik_{y}}\e^{in\pi\frac{3}{2}}\right)J_{n}\left(A_{0}\right),\\
h'_n & =
-J\left(\e^{-i\left(k_{y}\frac{1}{2}+k_{x}\frac{\sqrt{3}}{2}\right)}\e^{-in\frac{\pi}{6}}\right.\nonumber \\
 &\left. {}+\e^{-i\left(k_{y}\frac{1}{2}-k_{x}\frac{\sqrt{3}}{2}\right)}\e^{-in\frac{5}{6}\pi}
 +\e^{ik_{y}}\e^{in\frac{\pi}{2}}\right)J_{n}\left(A_{0}\right).
\end{align}

At small $\phi$, we enter the Landau-level regime. Because
of the suppression of the mixing with energy difference, the strongest 
overlap occurs between the highest and lowest Landau levels of two neighboring Floquet copies.  This observation justifies an
expansion of the Hamiltonian around the maximum of the spectrum
at $\mathbf{k}=0$. The dispersion is quadratic in leading order, and we find
\begin{align}
H_{0} &= -3JJ_{0}\left(A_{0}\right)\left[1-\frac{1}{4}\left(k_{x}^{2}+k_{y}^{2}\right)\right]\sigma_{x},\label{eq:-9}
\end{align}
where $\sigma_x$ is a Pauli matrix in the sublattice pseudospin space,
and we have omitted the higher order terms.
We now introduce the magnetic field by minimal Peierls substitution, and then
the standard ladder operators $a$ and $a^{\dagger}$ to find
\begin{align}
H_{0} & =-3JJ_{0}\left(A_{0}\right)\left[1-\frac{1}{2l_B^2}(a^{\dagger}a+\tfrac{1}{2})\right]\sigma_{x},\label{eq:-9-1}
\end{align}
where $l_B=\sqrt{\hbar/eB}$ is the magnetic length in terms of the magnetic field $B$. (We recall that the lattice spacing has been set to unity.)
The term $H_{1}$, mixing two copies of the butterfly spectrum, is
obtained by a similar calculation,
\begin{align}
H_{1} & =JJ_{1}\left(A_{0}\right)\frac{3}{2}\left[\frac{\sqrt{2}}{l_B}a^{\dagger}\sigma_{x}+\frac{1}{2l_B^2}aa\sigma_{y}\right].\label{eq:-11}
\end{align}
The eigenstates $\psi_{n,\pm}$ of $H_{0}$ have the same structure as the eigenstates of $\sigma_x$,
\begin{align}
\psi_{n,\pm} & =\frac{1}{\sqrt{2}}\left(\begin{array}{c}
|n\rangle\\
\mp|n\rangle
\end{array}\right),\label{eq:-15}
\end{align}
and their energies are
\begin{align}
E_{n,\pm} & =\pm3JJ_{0}\left(A_{0}\right)\left[1-\frac{1}{2\l_B^2}(n+\tfrac{1}{2})\right].\label{eq:-16}
\end{align}
These results are compatible with Ref.~\cite{quelle2016bandwidth}, which discusses the zero-field case.

One observes that for each Floquet copy, which we label by $r$ in the following, there are two sequences of Landau levels: one where
the zeroth Landau level is at the top of the spectrum, and one where
it is at the bottom of the spectrum, labeled by $+$ and
$-$, respectively. In $H_{1}$, the term proportional to $\sigma_y$
couples $\psi_{n,+,r}$ with $\psi_{n',-,r+1}$ and $\psi_{n,-,r}$ with
$\psi_{n',+,r+1}$.  The former pair constitutes states very close in energy
(energy difference $\Delta E \ll \hbar\omega$) whereas the latter pair are distant
states ($\Delta E \approx 2\hbar \omega$). The term proportional to $\sigma_x$ couples
$\psi_{n,\pm,r}$ with $\psi_{n',\pm,r+1}$, whose energy difference is
$\Delta E\approx \hbar\omega$. From perturbation theory, it follows that the energy
shift due to the mixing term scales as $1/\Delta E$. Consequently, hybridization
between the states $\psi_{n,+,r}$ and $\psi_{n',-,r+1}$ is significant, whereas
the couplings between the other pairs have negligible effects.

The strong mixing between $\psi_{n,+,r}$ and $\psi_{n',-,r+1}$ is due to the matrix
element proportional to $a a$ in Eq.~\eqref{eq:-11}.  Thus,
hybridization occurs between these states if $n' = n-2$.  In Figs.~\ref{fig:-1}(c) and \ref{fig:-1}(d),
we indeed observe that avoided crossings occur between the Landau levels labeled
0 (blue) and 2 (red), between 1 (blue) and 3 (red), etc. The top two
Landau levels of the lower copy ($n=0,1$, labeled in red) do not have a partner; 
they do not hybridize with any of the bands of the upper copy (labeled in blue).

\section{Numerical results\label{sec:Numerical-results}}

We now go beyond the low-flux regime and study the full Hofstadter butterfly.
We present numerical
results for both high frequencies, when the periodicity of the spectrum
is much larger than the bandwidth, and lower frequencies, where overlaps
are observed.

\begin{figure*}[t]
\includegraphics[width=9cm]{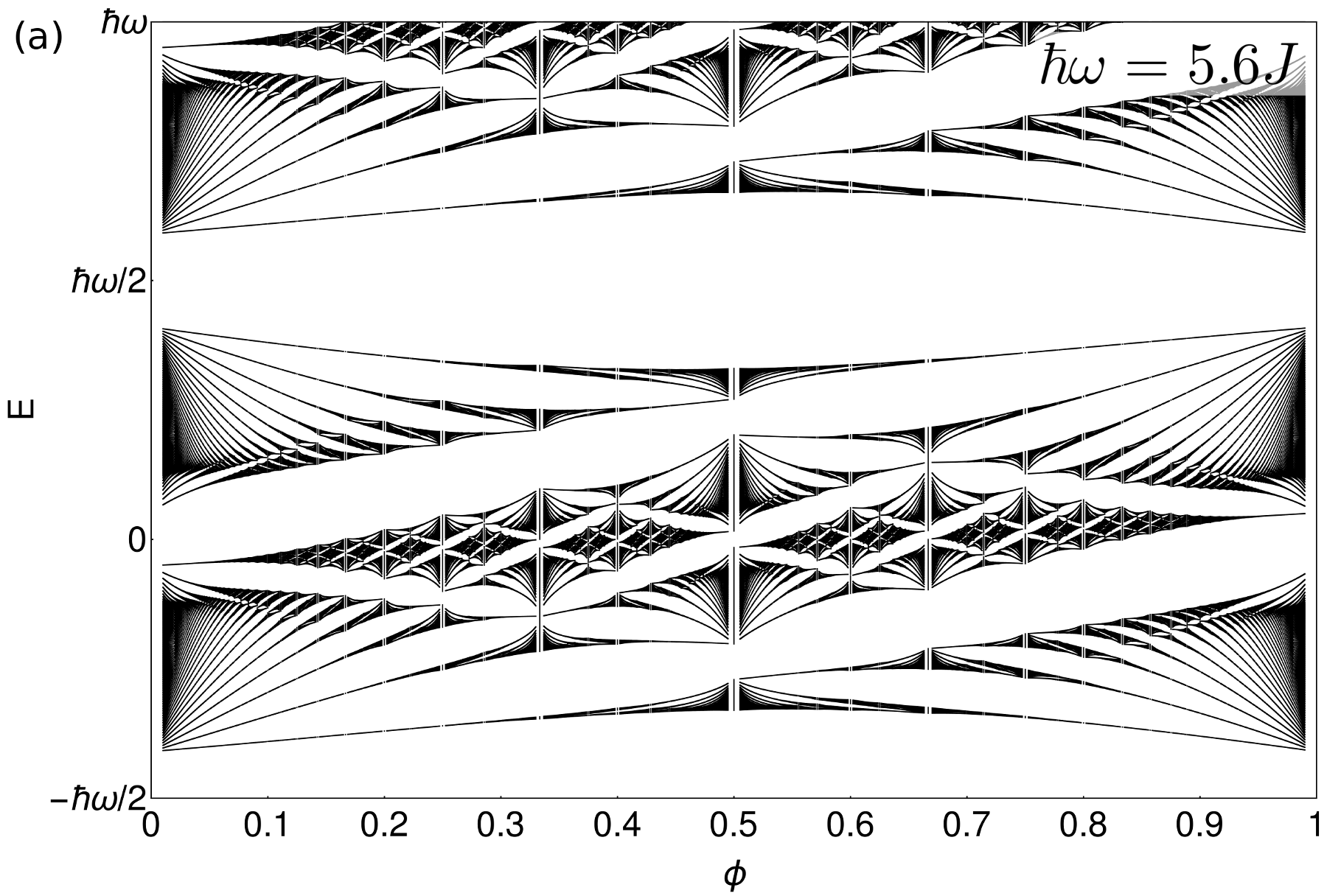}\includegraphics[width=9cm]{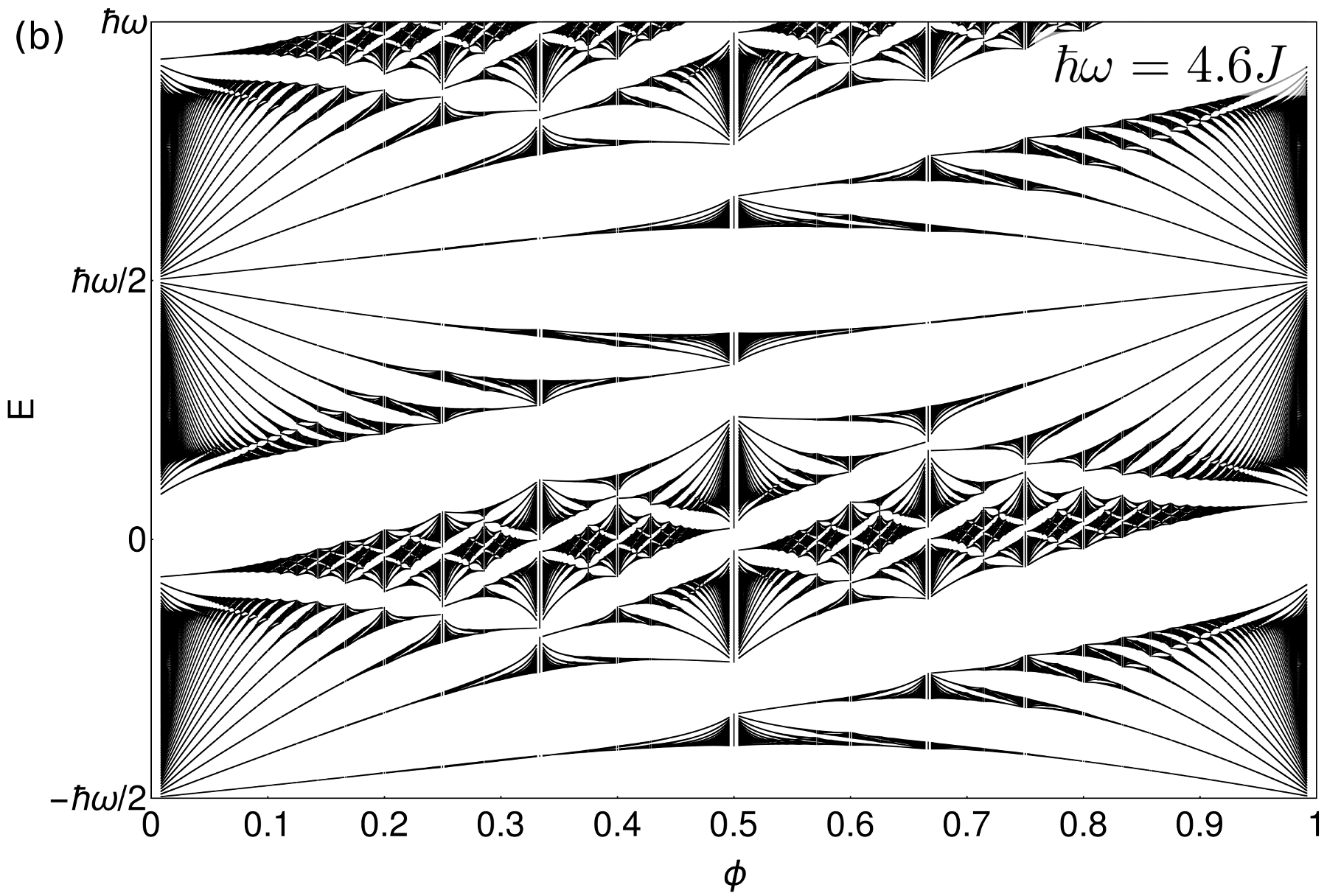}

\includegraphics[width=9cm]{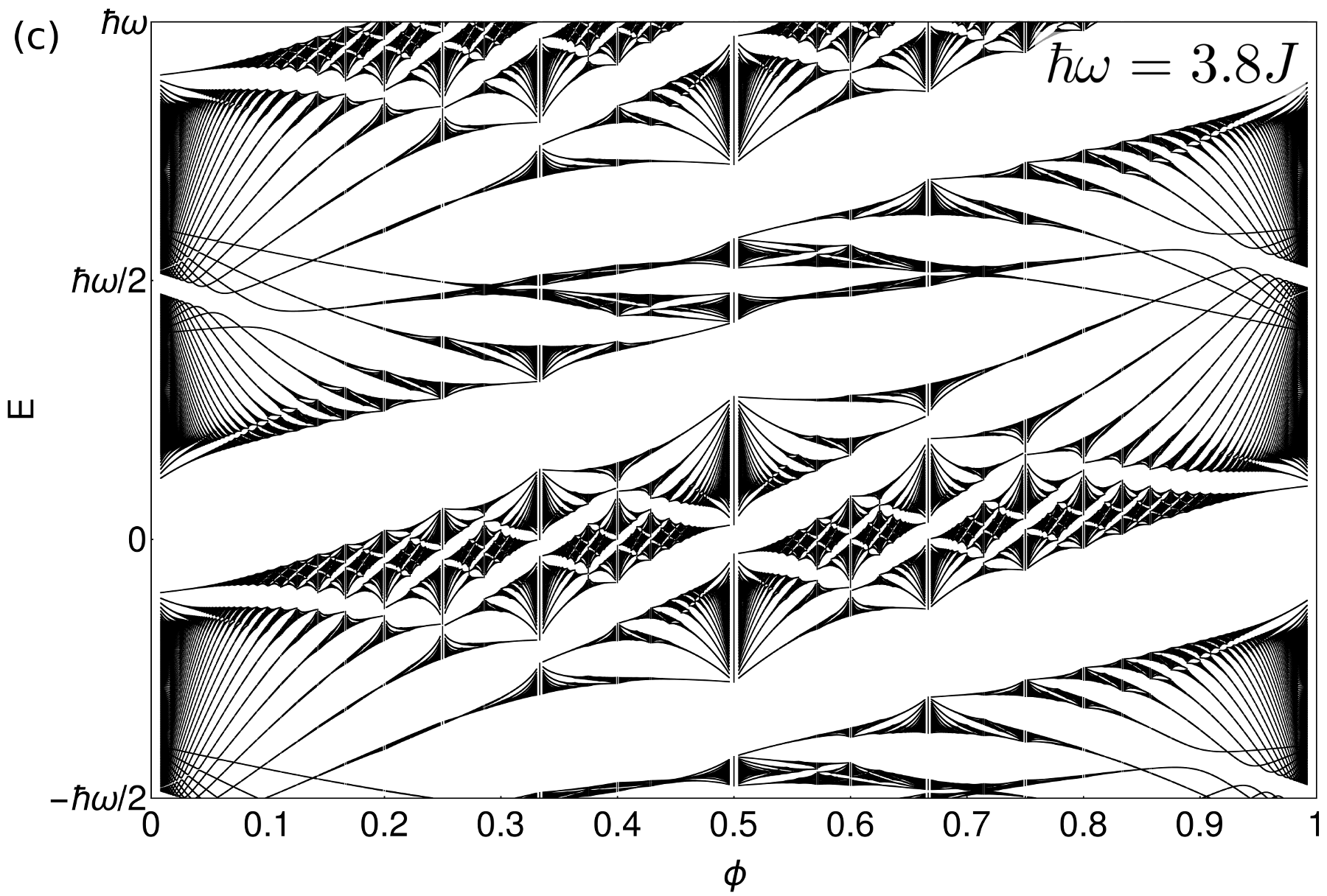}\includegraphics[width=9cm]{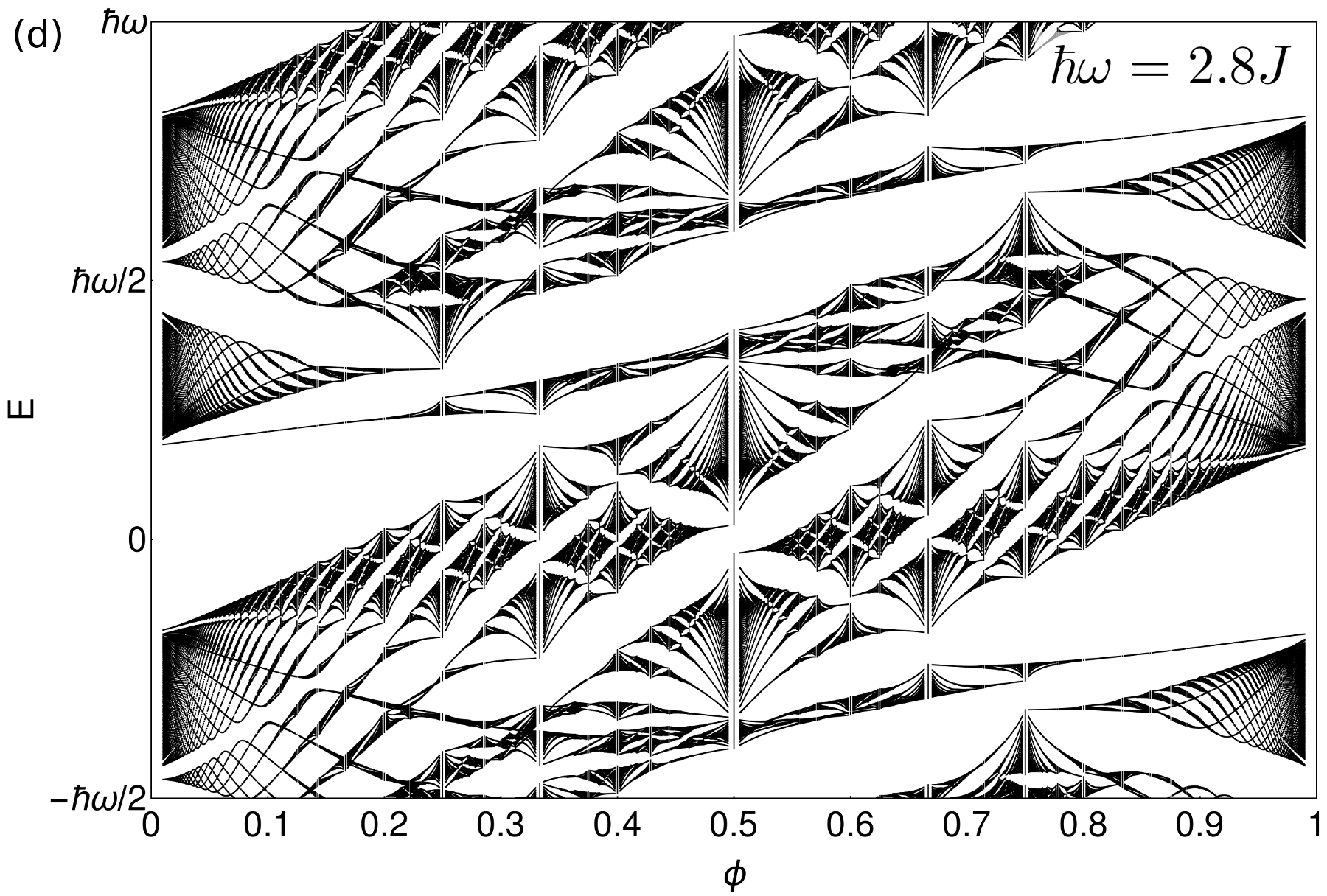}

\includegraphics[width=9cm]{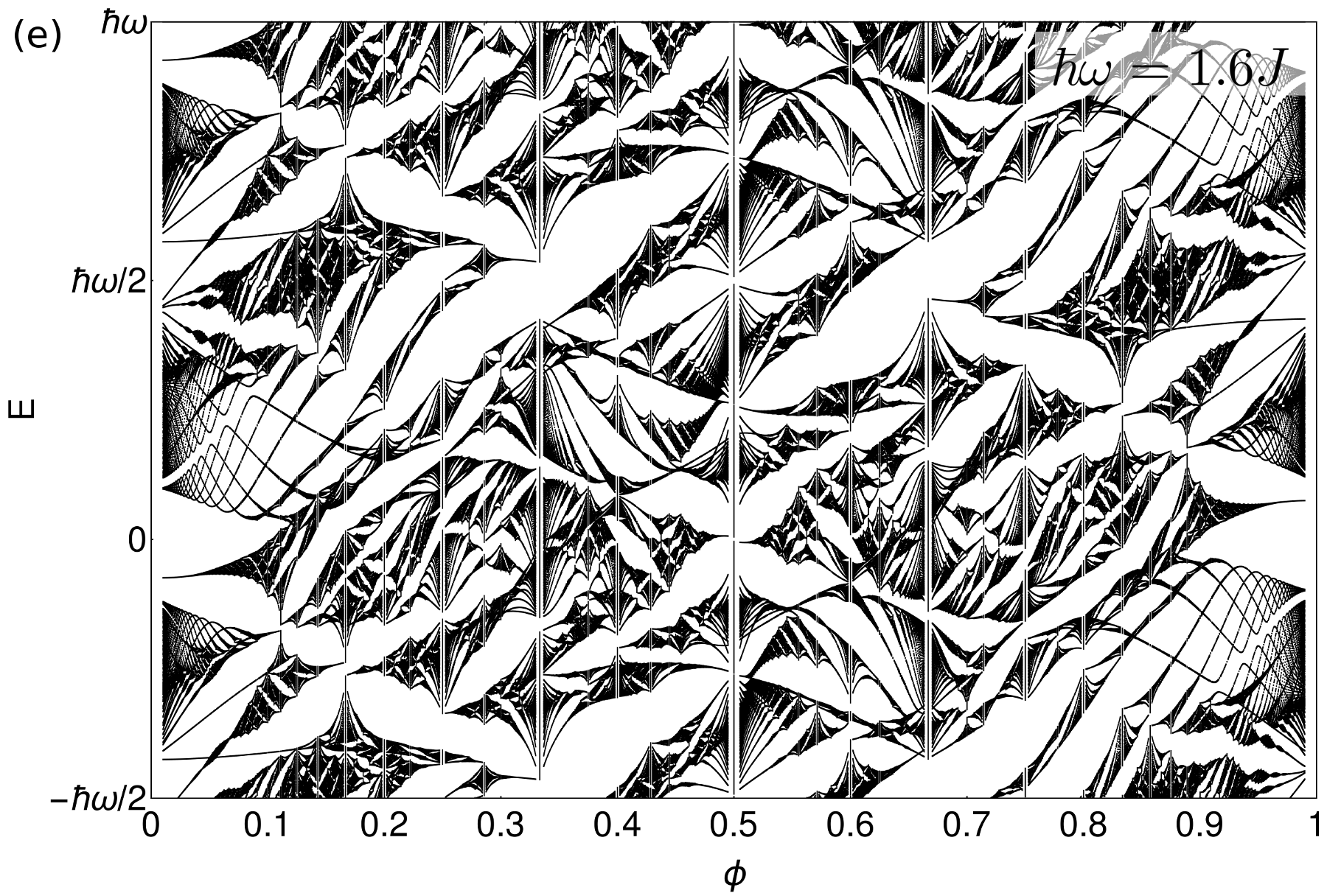}\includegraphics[width=9cm]{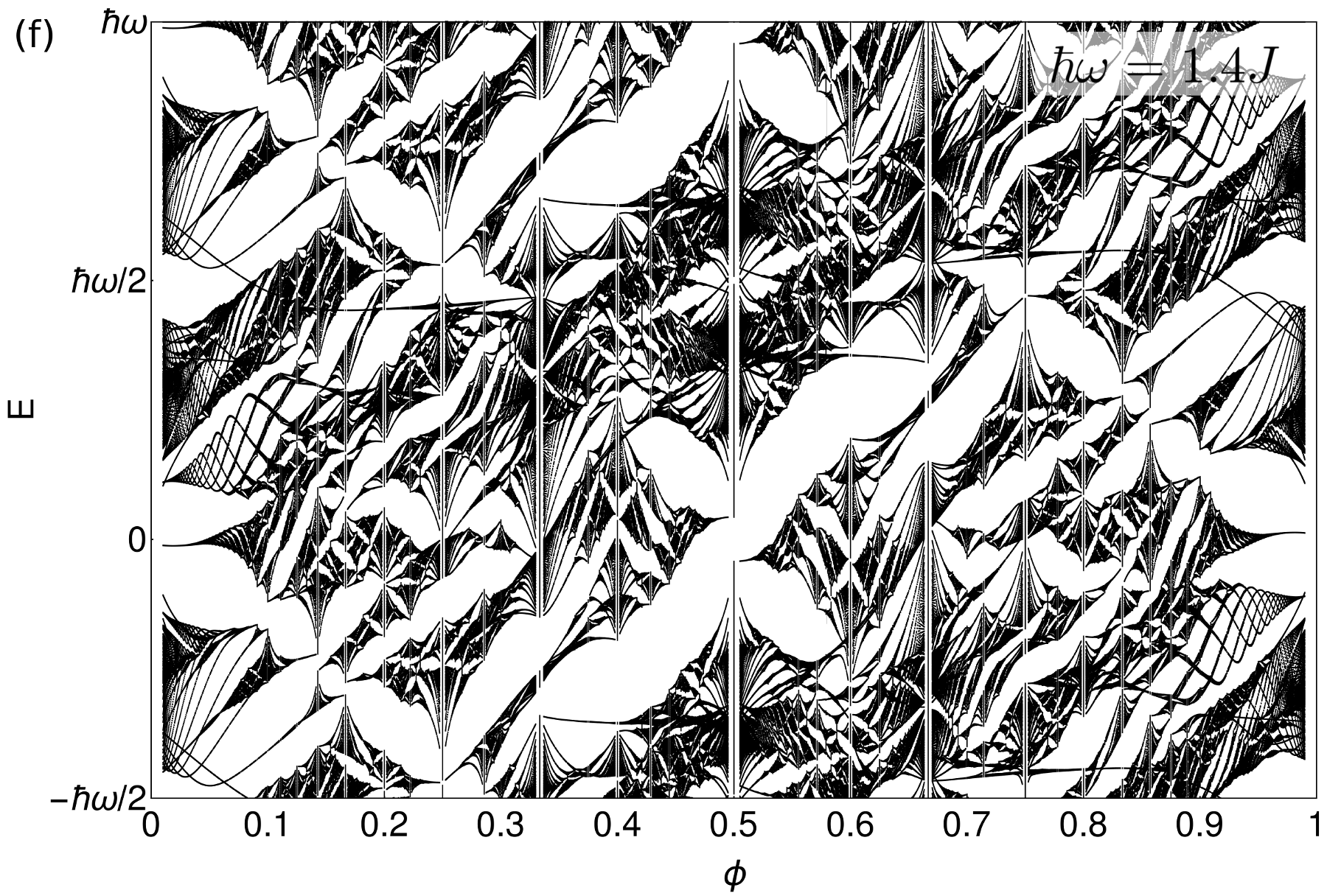}

\caption{Full spectra plotted for various values of the frequency and $A_{0}=1$.\label{fig:}}
\end{figure*}

\subsection*{High-frequency regime}

In Fig.~\ref{fig:Butterfly-spectrum-for}, we plot the energy levels
as a function of the flux per plaquette $\phi$, for several values
of the amplitude at a frequency of $\omega=12J/\hbar$, such that
the periodicity of the spectrum $\hbar\omega$ is larger than the
bandwidth $\sim6J$. Thus, there are only resonances within one Floquet
copy of the spectrum.

The colors of the gaps correspond to the associated topological invariants,
which are obtained by using the St\v{r}eda formula \cite{streda1982theory}
\begin{equation}
\sigma_\mathrm{H}=2\frac{e^{2}}{h}\frac{\partial N}{\partial\phi},\label{eq:-18}
\end{equation}
which provides the Hall conductivity $\sigma_\mathrm{H}$, in terms of the integrated
density of states $N$ and the conductance quantum $e^2/h$. We have checked and confirmed that the 
resulting values of $\sigma_\mathrm{H}$ from Eq.~\eqref{eq:-18}
are identical to those obtained
by counting the number of chiral edge states in a ribbon-geometry calculation
of the dispersion. Identical results can be obtained from explicit calculation of the Chern numbers \cite{fukui2005chern,wackerl2018driven}, however at a higher computational expense. Although the topological invariant
of Floquet systems is not the same as for static systems \cite{rudner2013anomalous},
in the high-frequency regime the St\v{r}eda formula still yields the correct
conductivity values because there is still a trivial gap between different
copies of the original spectrum.

For $\phi=0$ (no magnetic field), circularly polarized light opens
up a topological gap in the honeycomb system, and realizes a dynamical Haldane
model \cite{Haldane1988,oka2009photovoltaic,kitagawa2011transport,quelle2014dynamical,jotzu2014experimental}. Since the
spectrum is continuous as a function of $\phi$, this gap must persist
for non-zero $\phi$. From Fig.~\ref{fig:Butterfly-spectrum-for}, we see indeed
that it connects with the large gap above $E=0$, which also has topological
invariant $+1$. If we create a gap with opposite winding number (by
reversing the polarization of the light), the gap that opens up at
$\phi=0$ would connect to the lower large gap with invariant $-1$.
At other fractional fluxes, such as $\phi=1/2,\,1/3,\,1/4$, non-trivial gaps
also open at $E=0$ with the same chirality and with the topological invariant equal
to the denominator of the rational flux.

As we increase the amplitude of the light $A_{0}$, we change the effective
couplings $J J_{n}(A_{0})$ [see Eq.~\eqref{eq:-7}] in the Floquet Hamiltonian,
which induces additional topological phase transitions. These will happen by
the closing and opening of a gap that already exists without driving
\cite{BeugelingEA2012}. An example can be observed at $\phi=1/3$, where the
large gap around $E\approx-1$ with invariant $-1$ becomes smaller for
$A_{0}=1.3$, closes around $A_{0}\approx1.5$ and is reopened at $A_{0}=1.7$.

The gap closing occurs at three points in the Brillouin zone and the topological invariant changes from $-1$ to $+2$ (see
colors in Fig.~\ref{fig:Butterfly-spectrum-for}), consistent with the number of gap closing points. Because the system
still has magnetic-translation symmetry, the topological invariant
must satisfy the Diophantine equation \cite{dana1985quantised,kickedHarper}
\begin{equation}
p\,c+q\,d=1,\label{eq:-17}
\end{equation}
for flux $\phi=p/q$, where $c$ is the topological invariant and
$d$ is integer. This means that the topological invariant can only
change in multiples of $q$, which indeed agrees with our observation
at the third gap for flux $\phi=1/3$.

\subsection*{Higher photon resonances}

As we lower $\omega$, and $\hbar\omega$ becomes comparable to the bandwidth
($\lesssim6J$), bands from the next copy will start interacting with each
other (this regime in the case of $\phi=0$ has been studied in Ref.~\cite{MikamiEA2016}). We plot the spectrum for $A_{0}=1$ and various frequencies in
Fig.~\ref{fig:}. In Fig.~\ref{fig:}(a) ($\hbar\omega=5.6J$) there is still a
gap between the first and second copy of bands. In Fig.~\ref{fig:}(b), the
bands are at the verge of crossing, and in Fig.~\ref{fig:}(c)
($\hbar\omega\ll6J$) there is an overlap between the two copies. The mixing of
the energy bands gives rise to an intricate spectrum, and also causes many
topological phase transitions. One example is the gap that appears at
$E\approx\hbar\omega/2$ around $\phi=1/2$ in Fig.~\ref{fig:}(c). Since the
different copies are now starting to overlap, the periodicity of the spectrum
makes it difficult to define a reference value for the filling (integrated
density of states $N$) and the St\v{r}eda formula no longer \emph{a priori}
provides the correct topological invariant. As we decrease $\omega$ even
further, an almost flat band appears for small $\phi$ [see
Fig.~\ref{fig:}(d)], where the gap below (above) has Hall conductivity +1(+2). In this regime, it is possible to  clearly distinguish
between two gaps with a different number of edge states (the one above has two, the gap below one), where the gap above
the flat state has been created by the Floquet driving. This could facilitate
experiments, since the narrow and flat band persists for a wide range of flux.

\section{Conclusion\label{sec:Conclusion}}

By irradiating a honeycomb lattice subjected to a perpendicular magnetic field
with circularly polarized light, its Hofstadter butterfly exhibits an even richer
structure than its static counterpart. In particular, we can follow the formation of
wing-like structures in the spectrum at low flux and low frequencies.
The highest two Landau levels of the spectrum do not mix with the
overlapping copy, while the other levels do, as captured by our effective
analytical model.

To realize these features experimentally, the Floquet perturbation
and the flux per unit cell need to be large. The Floquet perturbations
enter through Bessel functions as factors of (reintroducing the lattice
constant $c$) $J_{n}\left(A_{0}c\right)$, which shows that a larger lattice
constant would increase the Floquet strength as well as the flux per
unit cell. This makes honeycomb structures with large lattice constants
a natural place to realize this system. Such structures can be for
example lattices of nanocrystals \cite{kalesaki2014dirac,beugeling2015topological} or optical
lattices \cite{soltan2011multi}. In optical lattices, one can also
implement shaking protocols \cite{zheng2014floquetShakingFloquet,goldman2016topological}. A circular shaking protocol will induce
a vector potential of the same form as Eq.~\eqref{eq:l} \cite{quelle2016bandwidth,quelle2017driving}.
The amplitude, however, will grow linearly with the frequency $\omega$,
while for light $A_{0}\sim eE/\hbar\omega$. As the required frequencies
are quite large, this will aid in an experimental realization. In
such a setup $\hbar\omega\approx2.7J$ can be realized, which would
be sufficient to observe the newly formed wings.

The structures observed at the process of opening the wings are
reminiscent of generic hybridized dispersions. For example, in
semiconductor quantum wells (e.g., HgCdTe/HgTe)
\cite{*[{Refer to Fig.~7.4 in }] [{}] PfeufferJeschke2000thesis,NovikEA2005,gomez2013floquetBlockexpansion},
gaps open between
Landau levels in the valence band. In that case, the "warping terms", which make
the dispersion non-isotropic, induce a coupling between Landau levels with
indices $n$ and $n\pm4$. The mechanism for the formation of these gaps is
thus analogous to the one governing the wing formation in the present Floquet model.
This analogy suggests a potential application of Floquet systems
as simulator of band structures of generic condensed matter systems.
In particular, such simulations could provide more insight into
hybridization in complicated Landau-level spectra.

\begin{acknowledgments}
We thank G. Platero for useful discussions. The work by A.Q. and C.M.S. is part of the D-ITP consortium, a program of the Netherlands Organisation for Scientific Research (NWO) that is funded by the Dutch Ministry of Education, Culture and Science (OCW). S.K. acknowledges support from a NWO-Graduate Program grant. 
\end{acknowledgments}

%

\end{document}